\documentclass[%
 reprint,
 twocolumn,
 amsmath,amssymb,
 aps,
 prl,
 nofootinbib,
 tightenlines,
 nobibnotes
]{revtex4-2}

\usepackage{graphicx}
\usepackage{dcolumn}
\usepackage{bm}
\usepackage{hyperref}

\usepackage{color}

\usepackage{gensymb}

\definecolor{Green}{RGB}{0,200,0}
\definecolor{Blue}{RGB}{51,153,255}
\definecolor{Red}{RGB}{151,010,010}

\begin{document}

\title{Experimental Measurement of Geometric Phase of Non-Geodesic Circles}

\author{Andrew A. Voitiv$^1$, Mark T. Lusk$^2$,}
\author{Mark E. Siemens$^1$}%
\email[]{msiemens@du.edu}%
\affiliation{$^1$Department of Physics and Astronomy, University of Denver, 2112 E. Wesley Avenue, Denver, CO 80208, USA
}%
\affiliation{
$^2$Department of Physics, Colorado School of Mines, 1500 Illinois Street, Golden, CO 80401, USA
}%

\date{\today}

\begin{abstract}
We present and implement a method for the experimental measurement of geometric phase of non-geodesic (small) circles on any SU(2) parameter space. This phase is measured by subtracting the dynamic phase contribution from the total phase accumulated. Our design does not require theoretical anticipation of this dynamic phase value and the methods are generally applicable to any system accessible to interferometric and projection measurements. Experimental implementations are presented for two settings: 1.) the Sphere of Modes of orbital angular momentum, and 2.) the Poincar\'e Sphere of polarizations of Gaussian beams.
\end{abstract}

\maketitle


Geometric phase plays an intrinsic role \cite{Berry2010GeometricMemories} in the evolution of all quantum and classical fields, from condensed matter to optics and beyond \cite{Cohen2019GeometricBeyond}. It has captured widespread interest both for applications such as electron transport in graphene \cite{Zhang2005ExperimentalGraphene}, braided photonic solid-state wave guides \cite{Noh2020BraidingModes}, and noise-resilient manipulation of solid-state spin-qubit quantum phases \cite{Yale2016OpticalQubit}, and also for its relevance at the intersection of geometry, information science, and quantum field theory \cite{wilczek1989geometric,Nayak2008Non-AbelianComputation, quigg2013gauge}. 

Optics, both classical and quantum, is a historically important and accessible arena for studying the geometric phase of state evolution. In fact, it was in an optical polarization setting that Pancharatnam first discovered a geometric phase~\cite{Pancharatnam1956GeneralizedPencils}, nearly thirty years prior to Berry's important identification of its quantum mechanical counterpart \cite{Berry1984QuantalChanges}. Geometric phase has since been given a holonomic foundation within differential geometry \cite{Simon1983HolonomyPhase}, subjected to theoretical examination in a variety of optical settings \cite{Martinelli1990ACircuits, Bhandari1991SU2Phases, Bhandari1991EvolutionDirection, Tiwari1992GeometricClassical, Bliokh2019GeometricAspects}, and experimentally realized in both classical (linear and nonlinear) and quantum optics \cite{Simon1988EvolvingExperiment, Bhandari1988ObservationInterferometer, Kwiat1991ObservationPhoton, Ortiz2014PolarimetricPhases,Maji2019GeometricBeams,Karnieli2019ExperimentalConversion,Hannonen2020MeasurementInterference}.

\begin{figure}[h!]
\centering
\includegraphics[width=\linewidth]{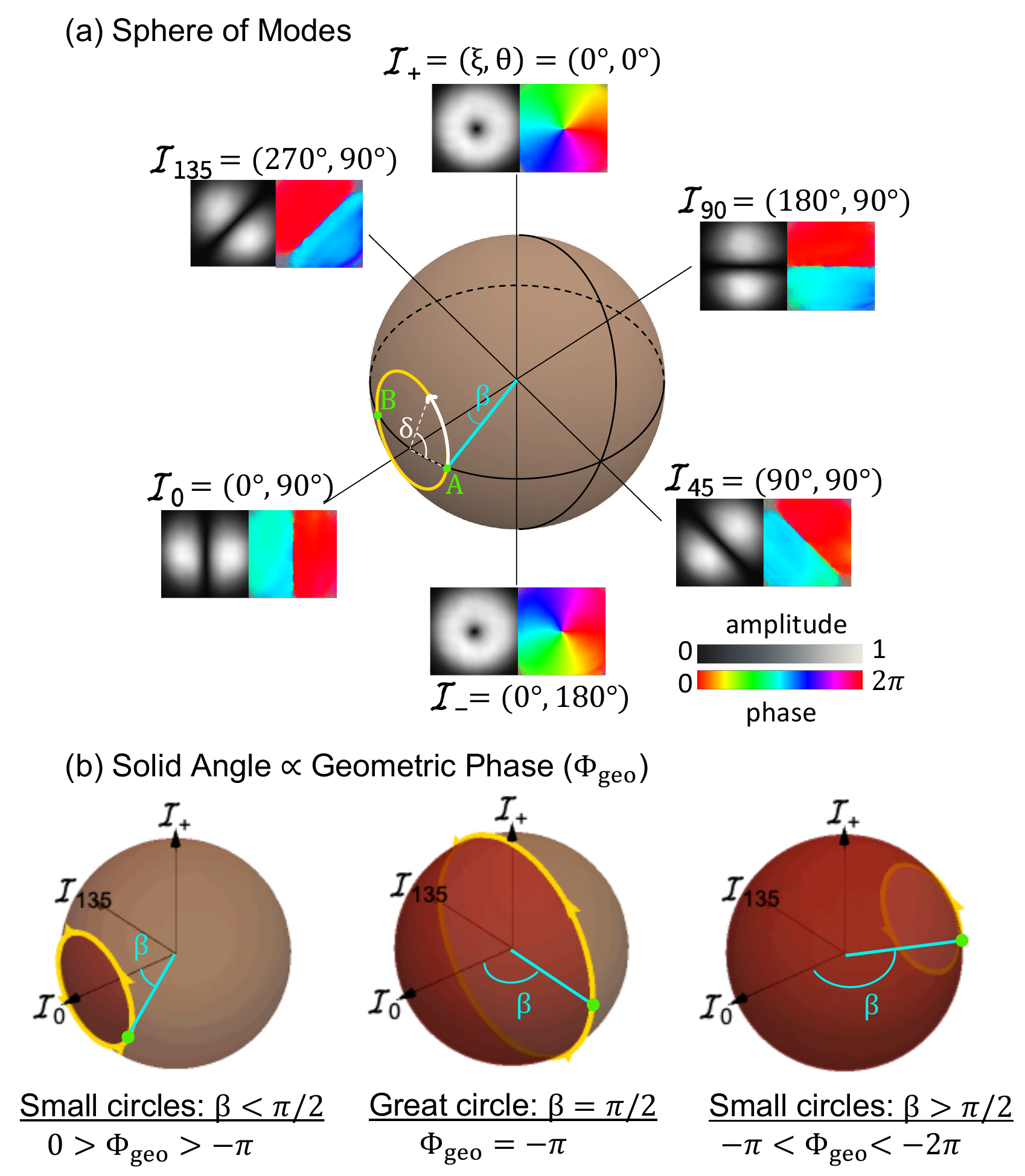}
\caption{\textbf{(a)} Sphere of Modes (SoM) for first-order Gaussian modes (optical vortices); similarly, the Poincar\'e Sphere (PS) exists for polarization with circular polarizations at the poles and linear polarizations along the equator  \cite{Bhandari1989SynthesisApproach}. Axes are labelled pictorially with experimental measurements of mode amplitude (grayscale) and phase (hue), along with spherical coordinates $(\xi, \theta)$. North and South poles are labeled with $\mathcal{I_{+}}$ and $\mathcal{I_{-}}$, respectively, for positively and negatively charged circular vortices. The equatorial axes are labeled with $\mathcal{I}_{0}$, $\mathcal{I}_{45}$, $\mathcal{I}_{90}$, and $\mathcal{I}_{135}$, corresponding with the tilt-angle (in degrees) of the Hermite-Gaussian modes. The yellow small circle is a circuit taken from a state $A$ at $(\xi,90\degree)$ to intermediate state $B$ at $(-\xi,90\degree)$ and then back to point $A$. \textbf{(b)} Depiction of how small circle geometric phase can exceed great circle geometric phase, depending on angle $\beta$ and according to $\Phi_{\mathrm{geo}} = -\Omega/2$, for solid angle $\Omega$ (colored red). Note that all trajectories trace their loops in the same direction \cite{Lusk2022TheLens}.}
\label{fig:ps}
\end{figure}

The investigation of geometric phase in association with the orbital angular momentum (OAM) of light \cite{vanEnk1993GeometricTransfer} emerged quickly after the explosion of work related to spin angular momentum (polarization). Fig. \ref{fig:ps} (a) shows a Sphere of Modes (SoM) for transformations of these modes, identified as the OAM analog of the polarization Poincar\'e Sphere (PS) \cite{SphereOfModes} which led to a series of demonstrations, over the past two decades, of  geometric phase accumulation for such optical vortices  \cite{Courtial1998MeasurementMomentum, Galvez2003GeometricMomentum, Habraken2010GeometricOrder, Alonso2017Ray-opticalBeams, Malhotra2018MeasuringInterferometry,Milione2012HigherLight, Voitiv2022TiltedGeodesics}. 
A recent contribution explicitly pinpoints the accumulation of this geometric phase with the evolution of beam waist, waist position, and fiber phase \cite{Cisowski2022Colloquium:Theory} which depend, in turn, on the transit fraction through optical elements \cite{Lusk2022TheLens}. It is fair to say that a theoretical understanding of the nature of geometric phase in classical paraxial optics is now well-established.

Fig. \ref{fig:ps} (b) illustrates the difference between small- and great-circle circuits and their enclosed solid angles. Geodesic arcs are contiguous sections of great circles---i.e., they are associated with circles of the same diameter as the sphere itself. Many experimental measurements of geometric phase associated with optical vortices or polarization have been careful to restrict attention to geodesic trajectories \cite{Galvez2003GeometricMomentum, Voitiv2022TiltedGeodesics}. The motivation for this is that the associated geometric phase, for an enclosed solid angle, is then equal to the total phase accumulation minus a standard propagation-dependent phase that can be removed using straightforward interferometry \cite{Bhandari1989SynthesisApproach, vanEnk1993GeometricTransfer,Galvez2003GeometricMomentum, Voitiv2022TiltedGeodesics}. Measuring the geometric phase of small circles is a greater challenge because it requires the removal of an additional contribution to dynamic phase \cite{Bhandari1989SynthesisApproach, Lusk2022TheLens}. For paradigms associated with evolving polarization, this additional dynamic phase has been removed either within particular experiments or with the aid of theoretical predictions. For example, it was quantified using intensity fringes for a two-pinhole interference experiment \cite{Hannonen2020MeasurementInterference}. A more general type of polarization experiment employed gauge transformations to negate all dynamic phase \cite{Ortiz2014PolarimetricPhases}; in that work, the dynamic phase was theoretically \textit{predicted} and then supplementary optics were used to remove it from the polarimetrically-measured total phase. Furthermore, there has been no purely experimental method for removing the non-geodesic contribution to dynamic phase that can be used with the OAM Sphere of Modes.

In this Letter, we show how the geometric phase of non-geodesic circular trajectories on a Sphere of Modes (SoM) or Poincar\'e sphere (PS) can be experimentally quantified. The key to our technique is the addition of a measurement halfway through the system evolution. The method is then implemented for both orbital and spin angular momentum transformations of Gaussian laser beams. While demonstrated within these specific optical settings, our simple theoretical framework is general and we expect the approach to be applicable to any system amenable to interferometric and projection measurements.


Consider the SoM shown in Fig. \ref{fig:ps}. Circuits on this sphere comprise a progression of vortex modes that can be realized with a combination of cylindrical lenses and Dove prisms \cite{Beijersbergen1993AstigmaticMomentum, vanEnk1993GeometricTransfer,SphereOfModes, Galvez2003GeometricMomentum,Lusk2022TheLens, Voitiv2022TiltedGeodesics}. Analogous circuits can be generated on the PS using waveplates to transform polarization~\cite{Bhandari1989SynthesisApproach, SphereOfModes}. As in \cite{Galvez2003GeometricMomentum} and \cite{Simon1988EvolvingExperiment}, we design our OAM and polarization experiments such that all phase measurements are made relative to a reference Gaussian beam. The dynamic phase associated with free-space propagation can therefore be removed and subsequent discussions of dynamic phase refer to phase accumulations due to non-geodesic paths on the PS or SoM. The geometric phase here does not rely on adiabatic dynamics \cite{Aharonov1987PhaseEvolution}.

The total, dynamic, and geometric phases of circular arcs (of arc length $\delta$) with opening angle $\beta$ on a PS are well-established theoretically \cite{Bhandari1989SynthesisApproach, Lusk2022TheLens}:
\begin{subequations} \label{subs}
\begin{equation} 
    \label{subtotal}
    \Phi_{\mathrm{tot}} = \mathrm{Arg}\langle A | B \rangle,
\end{equation}
\begin{equation} \label{subdynamic}
    \Phi_{\mathrm{dyn}} = -\frac{\delta}{2} \cos{\beta},
\end{equation}
\begin{equation} \label{subgeometric}
\begin{split}
    \Phi_{\mathrm{geo}} = \Phi_{\mathrm{tot}} - \Phi_{\mathrm{dyn}} .
\end{split}
\end{equation} 
\end{subequations}
Here $|A \rangle$ and $|B \rangle$ are spinors in SU(2) that correspond to the Stokes positions shown in Fig. \ref{fig:ps}. The geometric phase obtained is equivalent to the familiar representation as negative one-half the solid angle enclosed by the circuit, as depicted in Fig. \ref{fig:ps} (b).

A relationship can also be derived between the azimuthal angle subtended, $\delta$, and the total phase accumulation \cite{Lusk2022TheLens}:
\begin{equation} \label{delta}
    \Phi_{\mathrm{tot}} = -\arctan{\left( \cos{\frac{\delta}{2}} \, , \, \cos{\beta} \, \sin{\frac{\delta}{2}} \right)},
\end{equation}
where the sign-unambiguous arctangent is defined as $\arctan(x,y) = \arctan(y/x)$ with the quadrant accounted for.

For a complete small circle path on the PS or SoM ($\delta=2\pi$), Eqn. \ref{delta} yields $\Phi_\textrm{tot}=-\pi$ and Eqn. \ref{subdynamic} yields $\Phi_\textrm{dyn}=-\pi \cos \beta$. Experimental measurements of the states before and after a complete circular path (e.g. interferrometrically \cite{vanEnk1993GeometricTransfer,Galvez2003GeometricMomentum,Voitiv2022TiltedGeodesics}) can be plugged into Eqn. \ref{subtotal} to measure the total phase, but the dynamic phase is not accessible this way because path-taken information is lost using two states that coincide. The geometric phase is therefore inaccessible as well. This is the heart of the problem we address in this Letter and it warrants repeating for emphasis: one cannot measure the geometric phase of non-geodesic circles using only measurements of initial and final states because those two data points alone shed no information on the size of the circle---the $\cos{\beta}$ dependence. Therefore, we seek a method for measuring $\cos{\beta}$ for an arbitrary circular path on the PS, which would yield the geometric phase, Eqn. \ref{subgeometric}, from direct experimental measurements.

Our strategy is to determine $\Phi_\mathrm{dyn}$ by taking a measurement of the state after a half-circle trajectory. Such a trajectory can be easily obtained in optical systems by transmission through a Dove prism, a half waveplate (for the polarization PS) or a $\pi$-converter (for the vortex-mode SoM). A half-circle trajectory, $\delta=\pi$, has two important advantages: 1.) all of the phase components in Eqns. \ref{subs} are simply half of their values for the full circle, and 2.) Eqn. \ref{delta} yields $\Phi_\mathrm{tot} = -\pi/2$---independent of $\beta$. We will show that this permits measurements of the initial and  half-circle states to directly measure $\Phi_\mathrm{dyn}$. $\Phi_\mathrm{geo}$ can then be determined from Eqn. \ref{subgeometric}.

Two SU(2) spinors on the same circle, $|A \rangle$ and $|B \rangle$, can be written as
\begin{equation} \label{spinors}
    \begin{split}
        |A \rangle &= \cos{ \frac{\beta}{2} } |\leftrightarrow \, \rangle + \sin{ \frac{\beta}{2} } |\updownarrow \, \rangle, \\
        |B \rangle =& \cos{ \frac{\beta}{2} } \, e^{-i \delta/2} |\leftrightarrow \, \rangle + \sin{ \frac{\beta}{2} } \, e^{i \delta/2} |\updownarrow \, \rangle,
    \end{split}
\end{equation}
where $|\leftrightarrow \, \rangle$ and $|\updownarrow \, \rangle$ are general kets that correspond to the SoM or PS axes $\mathcal{I}_{0}$ and $\mathcal{I}_{90}$, respectively. There are equivalent orthogonal basis vectors for any SU(2) setting. Here, we use the Hermite Gaussian modes $|\mathrm{HG}_{10} \rangle$ and $|\mathrm{HG}_{01} \rangle$ for the SoM and linear polarization states $|\mathrm{H} \rangle$ and $|\mathrm{V} \rangle$ for the PS. This implies that
\begin{equation*}
    \langle A | B \rangle  = \cos^2{ \frac{\beta}{2} } \, e^{-i \delta/2} + \sin^2{ \frac{\beta}{2} } \, e^{i \delta/2}.
\end{equation*}
Collect up the real and imaginary parts to give
\begin{equation*}
    \begin{split}
        \langle A | B \rangle  = \left( \cos^2{ \frac{\beta}{2} } \cos{ \frac{\delta}{2} } + \sin^2{ \frac{\beta}{2} } \cos{ \frac{\delta}{2} } \right) \\
        + \, \, i \, \left( \sin^2{ \frac{\beta}{2} } \sin{\frac{\delta}{2} } - \cos^2{ \frac{\beta}{2} } \sin{ \frac{\delta}{2} } \right).
    \end{split}
\end{equation*}
This can be immediately simplified to
\begin{equation} 
    \langle A | B \rangle = \cos{ \frac{\delta}{2} } - i \, \cos{\beta}\sin{\frac{\delta}{2} },
\end{equation}
and taking the imaginary part yields
\begin{equation} \label{beta}
    \cos{\beta} = -\csc{ \frac{\delta}{2} } \, \mathrm{Im} \langle A | B \rangle .
\end{equation}

In the case of a half-circle path, $\delta=\pi$, and the result simplifies to
\begin{equation}\label{innerproduct}
    \cos{\beta} = -\mathrm{Im}\langle A_0 | B_{\pi} \rangle,
\end{equation}
for initial state $|A_0 \rangle (\delta=0)$ and intermediate state $|B_{\pi} \rangle (\delta=\pi)$ connected by a half-circle.

A reformulation of the total, dynamic, and geometric phase terms is now possible in terms of experimentally-measurable quantities:
\begin{subequations} \label{main}
\begin{equation} 
    \label{subtotalEXP}
    \Phi_{\mathrm{tot}} = \mathrm{Arg} \langle A_0 | C_{2\pi} \rangle,
\end{equation}
\begin{equation} \label{subdynamicEXP}
    \Phi_{\mathrm{dyn}} = \mathrm{Arg} \langle A_0 | C_{2\pi} \rangle \, \mathrm{Im}\langle A_0 | B_{\pi} \rangle,
\end{equation}
\begin{equation} \label{subgeometricEXP}
\begin{split}
    \Phi_{\mathrm{geo}} = \Phi_{\mathrm{tot}} - \Phi_{\mathrm{dyn}} = \mathrm{Arg}\langle A_0 | C_{2\pi} \rangle \, \left(1 - \mathrm{Im} \langle A_0 | B_{\pi} \rangle \right) ,
\end{split}
\end{equation} 
\end{subequations}
for $|C_{2\pi}\rangle (\delta=2\pi$) being the final state that completes the circle.

\begin{figure}[h!]
\centering
\includegraphics[width=\linewidth]{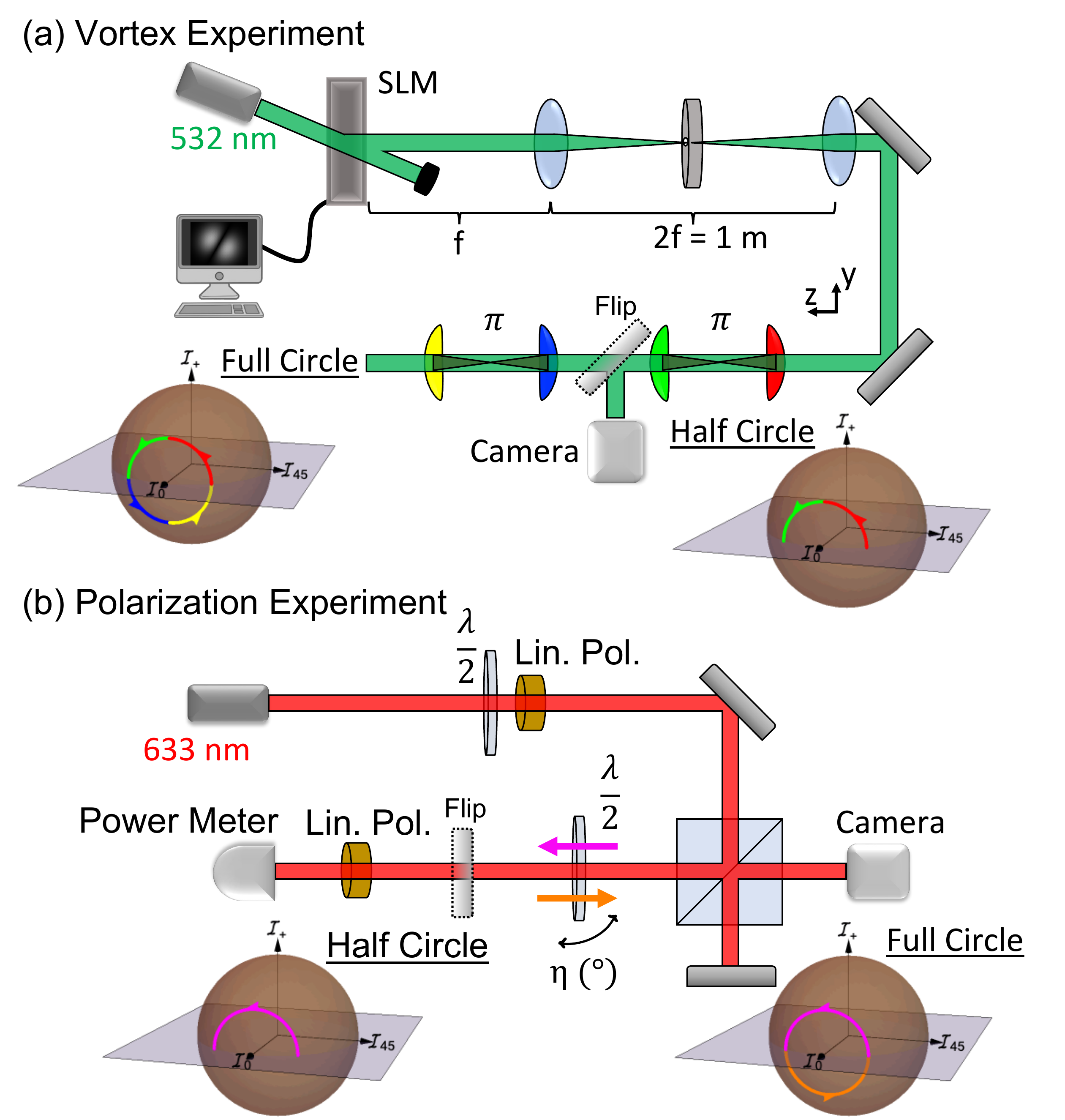}
\caption{\textbf{(a)} In the vortex experiments, a collimated, single-mode Gaussian passes through a hologram on a transmissive spatial light modulator (SLM), from which the first diffracted order is imaged onto a series of $\pi$-converters (labelled ``$\pi$''). The vortex beam is measured before the first $\pi$-converter (initial state $|A \rangle$) and then after (intermediate state $|B \rangle$), which makes a half circle trajectory. Going through the second $\pi$-converter, the final state of the beam coincides with the initial state after making a complete circle on the SoM. These circuits are analytically calculated and plotted \cite{Lusk2022TheLens}. \textbf{(b)} In the polarization experiments, a collimated, single-mode Gaussian is locked to a horizontal polarization with the first linear polarizer (``Lin. Pol.''). At the beamsplitter, one arm is unmodified to be used to interfere with the other arm after it passes through the half-waveplate twice (flip mirror is up)---this produces a full circle circuit, from which the total phase is measured. Dynamic phase is measured after one transit through the waveplate (flip mirror is down) with a linear polarizer, locked to the same polarization as the first ``Lin. Pol.'' The orientation of the waveplate, $\eta$, is rotated to increase $\beta$ on the PS.}
\label{fig:schematic}
\end{figure}

We demonstrate the implementation of this algorithm for measuring geometric phase for two different optical bases. Our first implementation uses vortex modes and a $\pi$-converter to provide the needed half-circle trajectory, as shown in the schematic of Fig. \ref{fig:schematic} (a). A collimated, single-mode Gaussian is transmitted through a spatial light modulator (SLM) \cite{Huang2012ALabs}, which crafts an initial vortex state, $|A \rangle$. Details on how these vortices are made experimentally are provided in the Supplementary Information. $|A \rangle$ is completely characterized by a camera before the first $\pi$-converter; that is, we measure both amplitude (square-root of the recorded intensity) and phase (measured via phase-shifting digital holography \cite{Andersen2019CharacterizingHolography}). We then fit this complex field to an explicit expression for the input beam in which a tunable initial phase-shift, $\phi$, is included to account for shifts associated with propagation through the apparatus that produces the beam.

The beam in state $|A \rangle$ is then propagated through the first $\pi$-converter. Between the mode converters, we measure intermediate state $|B \rangle$ and characterize the mode using the same methods used to characterize state $|A \rangle$. With both fields now distilled into discrete data arrays, we can immediately evaluate the total phase of Eqn. \ref{subtotal} and the inner product of  Eqn. \ref{innerproduct}. To complete the circle, the beam then transits an identical, second $\pi$-converter. However, rather than performing a second measurement, we simply scale the intermediate-measured phases by two, as discussed above.

\begin{figure}[h!]
\centering
\includegraphics[width=.96\linewidth]{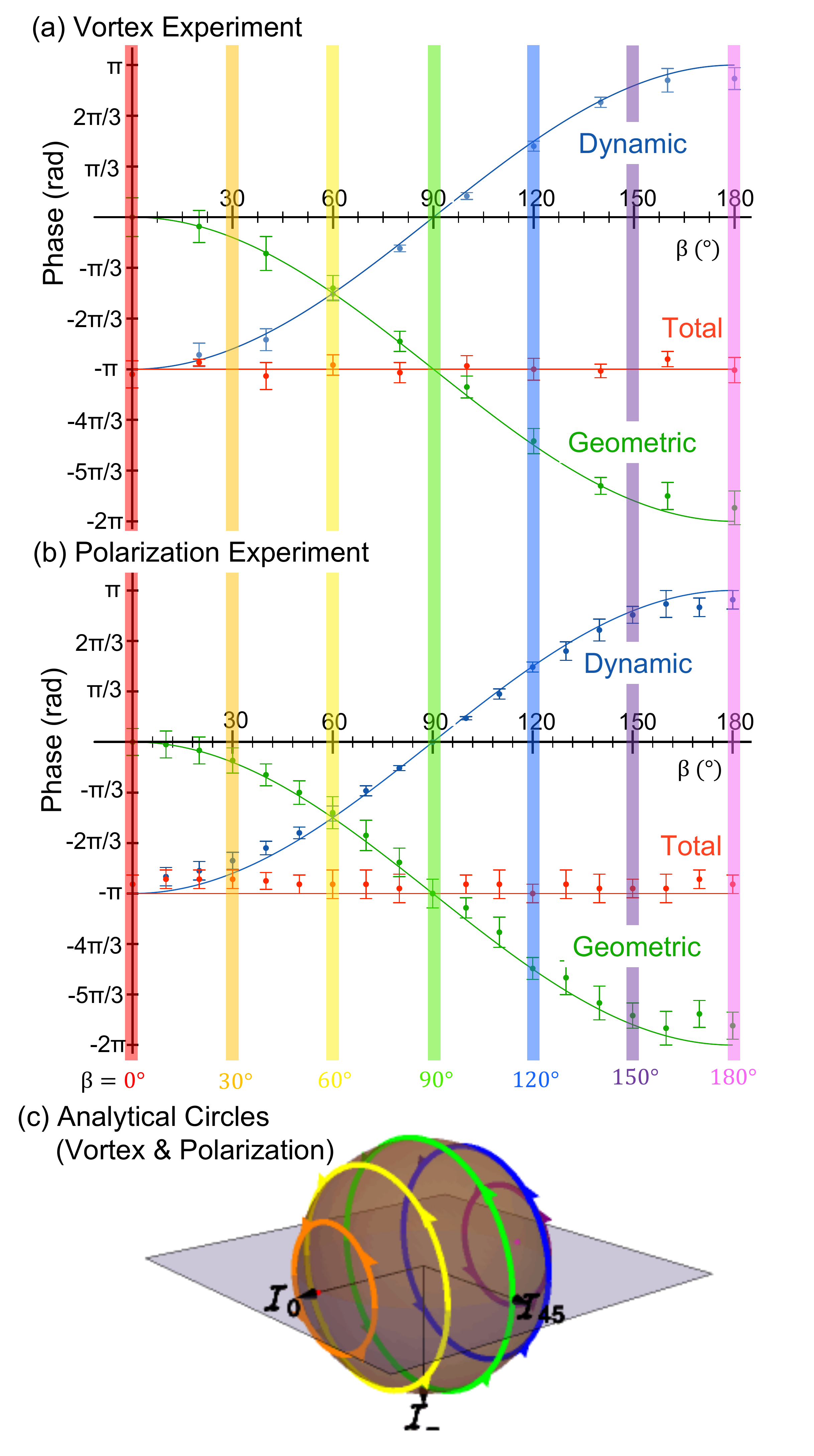}
\caption{\textbf{(a)} Dynamic (blue), total (red), and geometric (green) phases of vortex transformations driven by a series of two $\pi$-converters, as a function of $\beta$. Dots with error bars are experimental data (average and standard deviation of five measurements) and curves are calculated (not fit to data) from Eqns. \ref{subtotal}, \ref{subdynamic}, and \ref{subgeometric}. \textbf{(b)} Similar phase measurements in the polarization domain, plotted with the same methods. Here, the waveplate was rotated up to $\eta = 90\degree$, corresponding to $\beta = 0\degree \rightarrow 180\degree$. \textbf{(c)} Sample analytical trajectories for the corresponding values of $\beta$ used in the data above, being functionally equivalent for the two cases \cite{Lusk2022TheLens}.}
\label{fig:results}
\end{figure}

This procedure is applied for a range of initial states with the results presented in Fig. \ref{fig:results} (a). The solid curves in the figure are theoretical predictions obtained from Eqns. \ref{subtotal} (total), \ref{subdynamic} (dynamic), and \ref{subgeometric} (geometric). The match of the measurement to theory (without fitting parameters) is striking. A detailed, step-by-step example is provided in the Supplementary Information for a representative measurement.

Our second implementation of the new methodology is associated with spin angular momentum (polarization), as shown in Fig. \ref{fig:schematic} (b). We fix the input state, $|A \rangle$, as horizontally polarized, and pass it through a rotatable half-waveplate, the polarization analog of a $\pi$-converter for OAM. Fixing the initial state and rotating the waveplate amounts to a change in the frame of reference used for OAM, where it is the initial state that is changed and the $\pi$-converter that is fixed. However, it is only the relative motion that is relevant. 

To measure the total phase accumulated over a half-circular trajectory, the state emerging from a double transmission through the waveplate is interfered with the initial state $|A \rangle$. The phase shift, $\Phi_{\mathrm{tot}}$, does not vary with waveplate rotation as shown in red in Fig. \ref{fig:results} (b). To measure the dynamic phase, we take the intermediate state, $|B \rangle$, after one pass through the half-waveplate and transmit it through a linear polarizer that is locked to the initial polarization of $|A \rangle$. This is a projection measurement (Eqn. \ref{innerproduct}). The analysis then proceeds as for OAM but with one caveat; the polarizer acts on field intensity rather than the field itself. (See the Supplementary Information for details on converting this measured Malus' law to the dynamic phase.) A complete set of results are shown in Fig. \ref{fig:results} (b) and, once again, is found to be in strong agreement with the theoretical predictions shown as solid curves.


In conclusion, this Letter presents a general method for measuring the geometric phase for small circle circuits in an SU(2) parameter space, based on mode measurements and projections of the initial and half-circle states. The approach is implemented for the first experimental measurements of small circle geometric phase for both orbital and spin angular momentum transformations of Gaussian laser beams. The derivation associated with Eqns. \ref{spinors} through  \ref{main} is not restricted to the two experimental settings we have performed here, which allows for our approach to be easily adapted. We anticipate that the results of this paper will find application to a variety of settings for which geometric phase has not yet been measured in association with small circles.

The authors acknowledge R. N. Mundell for assisting on the polarization experiment.

\noindent \textbf{Funding.} W. M. Keck Foundation; NSF (1553905)

\noindent \textbf{Disclosures.} The authors declare no conflicts of interest.

\noindent \textbf{Data availability.} Data underlying all results presented are available from the authors upon reasonable request.

\bibliography{Refs}

\begin{thebibliography}{40}%
\makeatletter
\providecommand \@ifxundefined [1]{%
 \@ifx{#1\undefined}
}%
\providecommand \@ifnum [1]{%
 \ifnum #1\expandafter \@firstoftwo
 \else \expandafter \@secondoftwo
 \fi
}%
\providecommand \@ifx [1]{%
 \ifx #1\expandafter \@firstoftwo
 \else \expandafter \@secondoftwo
 \fi
}%
\providecommand \natexlab [1]{#1}%
\providecommand \enquote  [1]{``#1''}%
\providecommand \bibnamefont  [1]{#1}%
\providecommand \bibfnamefont [1]{#1}%
\providecommand \citenamefont [1]{#1}%
\providecommand \href@noop [0]{\@secondoftwo}%
\providecommand \href [0]{\begingroup \@sanitize@url \@href}%
\providecommand \@href[1]{\@@startlink{#1}\@@href}%
\providecommand \@@href[1]{\endgroup#1\@@endlink}%
\providecommand \@sanitize@url [0]{\catcode `\\12\catcode `\$12\catcode
  `\&12\catcode `\#12\catcode `\^12\catcode `\_12\catcode `\%12\relax}%
\providecommand \@@startlink[1]{}%
\providecommand \@@endlink[0]{}%
\providecommand \url  [0]{\begingroup\@sanitize@url \@url }%
\providecommand \@url [1]{\endgroup\@href {#1}{\urlprefix }}%
\providecommand \urlprefix  [0]{URL }%
\providecommand \Eprint [0]{\href }%
\providecommand \doibase [0]{https://doi.org/}%
\providecommand \selectlanguage [0]{\@gobble}%
\providecommand \bibinfo  [0]{\@secondoftwo}%
\providecommand \bibfield  [0]{\@secondoftwo}%
\providecommand \translation [1]{[#1]}%
\providecommand \BibitemOpen [0]{}%
\providecommand \bibitemStop [0]{}%
\providecommand \bibitemNoStop [0]{.\EOS\space}%
\providecommand \EOS [0]{\spacefactor3000\relax}%
\providecommand \BibitemShut  [1]{\csname bibitem#1\endcsname}%
\let\auto@bib@innerbib\@empty
\bibitem [{\citenamefont {Berry}(2010)}]{Berry2010GeometricMemories}%
  \BibitemOpen
  \bibfield  {author} {\bibinfo {author} {\bibfnamefont {M.}~\bibnamefont
  {Berry}},\ }\bibfield  {title} {\bibinfo {title} {{Geometric phase
  memories}},\ }\href {https://doi.org/10.1038/nphys1608} {\bibfield  {journal}
  {\bibinfo  {journal} {Nature Physics}\ }\textbf {\bibinfo {volume} {6}},\
  \bibinfo {pages} {148} (\bibinfo {year} {2010})}\BibitemShut {NoStop}%
\bibitem [{\citenamefont {Cohen}\ \emph {et~al.}(2019)\citenamefont {Cohen},
  \citenamefont {Larocque}, \citenamefont {Bouchard}, \citenamefont
  {Nejadsattari}, \citenamefont {Gefen},\ and\ \citenamefont
  {Karimi}}]{Cohen2019GeometricBeyond}%
  \BibitemOpen
  \bibfield  {author} {\bibinfo {author} {\bibfnamefont {E.}~\bibnamefont
  {Cohen}}, \bibinfo {author} {\bibfnamefont {H.}~\bibnamefont {Larocque}},
  \bibinfo {author} {\bibfnamefont {F.}~\bibnamefont {Bouchard}}, \bibinfo
  {author} {\bibfnamefont {F.}~\bibnamefont {Nejadsattari}}, \bibinfo {author}
  {\bibfnamefont {Y.}~\bibnamefont {Gefen}},\ and\ \bibinfo {author}
  {\bibfnamefont {E.}~\bibnamefont {Karimi}},\ }\bibfield  {title} {\bibinfo
  {title} {{Geometric phase from Aharonov–Bohm to Pancharatnam–Berry and
  beyond}},\ }\href {https://doi.org/10.1038/s42254-019-0071-1} {\bibfield
  {journal} {\bibinfo  {journal} {Nature Reviews Physics}\ }\textbf {\bibinfo
  {volume} {1}},\ \bibinfo {pages} {437} (\bibinfo {year} {2019})}\BibitemShut
  {NoStop}%
\bibitem [{\citenamefont {Zhang}\ \emph {et~al.}(2005)\citenamefont {Zhang},
  \citenamefont {Tan}, \citenamefont {Stormer},\ and\ \citenamefont
  {Kim}}]{Zhang2005ExperimentalGraphene}%
  \BibitemOpen
  \bibfield  {author} {\bibinfo {author} {\bibfnamefont {Y.}~\bibnamefont
  {Zhang}}, \bibinfo {author} {\bibfnamefont {Y.~W.}\ \bibnamefont {Tan}},
  \bibinfo {author} {\bibfnamefont {H.~L.}\ \bibnamefont {Stormer}},\ and\
  \bibinfo {author} {\bibfnamefont {P.}~\bibnamefont {Kim}},\ }\bibfield
  {title} {\bibinfo {title} {{Experimental observation of the quantum Hall
  effect and Berry's phase in graphene}},\ }\href
  {https://doi.org/10.1038/nature04235} {\bibfield  {journal} {\bibinfo
  {journal} {Nature}\ }\textbf {\bibinfo {volume} {438}},\ \bibinfo {pages}
  {201} (\bibinfo {year} {2005})}\BibitemShut {NoStop}%
\bibitem [{\citenamefont {Noh}\ \emph {et~al.}(2020)\citenamefont {Noh},
  \citenamefont {Schuster}, \citenamefont {Iadecola}, \citenamefont {Huang},
  \citenamefont {Wang}, \citenamefont {Chen}, \citenamefont {Chamon},\ and\
  \citenamefont {Rechtsman}}]{Noh2020BraidingModes}%
  \BibitemOpen
  \bibfield  {author} {\bibinfo {author} {\bibfnamefont {J.}~\bibnamefont
  {Noh}}, \bibinfo {author} {\bibfnamefont {T.}~\bibnamefont {Schuster}},
  \bibinfo {author} {\bibfnamefont {T.}~\bibnamefont {Iadecola}}, \bibinfo
  {author} {\bibfnamefont {S.}~\bibnamefont {Huang}}, \bibinfo {author}
  {\bibfnamefont {M.}~\bibnamefont {Wang}}, \bibinfo {author} {\bibfnamefont
  {K.~P.}\ \bibnamefont {Chen}}, \bibinfo {author} {\bibfnamefont
  {C.}~\bibnamefont {Chamon}},\ and\ \bibinfo {author} {\bibfnamefont {M.~C.}\
  \bibnamefont {Rechtsman}},\ }\bibfield  {title} {\bibinfo {title} {{Braiding
  photonic topological zero modes}},\ }\href
  {https://doi.org/10.1038/s41567-020-1007-5} {\bibfield  {journal} {\bibinfo
  {journal} {Nature Physics}\ }\textbf {\bibinfo {volume} {16}},\ \bibinfo
  {pages} {989} (\bibinfo {year} {2020})}\BibitemShut {NoStop}%
\bibitem [{\citenamefont {Yale}\ \emph {et~al.}(2016)\citenamefont {Yale},
  \citenamefont {Heremans}, \citenamefont {Zhou}, \citenamefont {Auer},
  \citenamefont {Burkard},\ and\ \citenamefont
  {Awschalom}}]{Yale2016OpticalQubit}%
  \BibitemOpen
  \bibfield  {author} {\bibinfo {author} {\bibfnamefont {C.~G.}\ \bibnamefont
  {Yale}}, \bibinfo {author} {\bibfnamefont {F.~J.}\ \bibnamefont {Heremans}},
  \bibinfo {author} {\bibfnamefont {B.~B.}\ \bibnamefont {Zhou}}, \bibinfo
  {author} {\bibfnamefont {A.}~\bibnamefont {Auer}}, \bibinfo {author}
  {\bibfnamefont {G.}~\bibnamefont {Burkard}},\ and\ \bibinfo {author}
  {\bibfnamefont {D.~D.}\ \bibnamefont {Awschalom}},\ }\bibfield  {title}
  {\bibinfo {title} {{Optical manipulation of Berry phase in a solid-state spin
  qubit}},\ }\href@noop {} {\bibfield  {journal} {\bibinfo  {journal} {Nature
  Photonics}\ }\textbf {\bibinfo {volume} {10}},\ \bibinfo {pages} {184}
  (\bibinfo {year} {2016})}\BibitemShut {NoStop}%
\bibitem [{\citenamefont {Wilczek}\ and\ \citenamefont
  {Shapere}(1989)}]{wilczek1989geometric}%
  \BibitemOpen
  \bibfield  {author} {\bibinfo {author} {\bibfnamefont {F.}~\bibnamefont
  {Wilczek}}\ and\ \bibinfo {author} {\bibfnamefont {A.}~\bibnamefont
  {Shapere}},\ }\href {https://books.google.com/books?id=5jOvlny96AkC} {\emph
  {\bibinfo {title} {{Geometric Phases in Physics}}}},\ Advanced series in
  mathematical physics\ (\bibinfo  {publisher} {World Scientific},\ \bibinfo
  {year} {1989})\BibitemShut {NoStop}%
\bibitem [{\citenamefont {Nayak}\ \emph {et~al.}(2008)\citenamefont {Nayak},
  \citenamefont {Simon}, \citenamefont {Stern}, \citenamefont {Freedman},\ and\
  \citenamefont {Das~Sarma}}]{Nayak2008Non-AbelianComputation}%
  \BibitemOpen
  \bibfield  {author} {\bibinfo {author} {\bibfnamefont {C.}~\bibnamefont
  {Nayak}}, \bibinfo {author} {\bibfnamefont {S.~H.}\ \bibnamefont {Simon}},
  \bibinfo {author} {\bibfnamefont {A.}~\bibnamefont {Stern}}, \bibinfo
  {author} {\bibfnamefont {M.}~\bibnamefont {Freedman}},\ and\ \bibinfo
  {author} {\bibfnamefont {S.}~\bibnamefont {Das~Sarma}},\ }\bibfield  {title}
  {\bibinfo {title} {{Non-Abelian anyons and topological quantum
  computation}},\ }\href {https://doi.org/10.1103/RevModPhys.80.1083}
  {\bibfield  {journal} {\bibinfo  {journal} {Reviews of Modern Physics}\
  }\textbf {\bibinfo {volume} {80}},\ \bibinfo {pages} {1083} (\bibinfo {year}
  {2008})}\BibitemShut {NoStop}%
\bibitem [{\citenamefont {Quigg}(2013)}]{quigg2013gauge}%
  \BibitemOpen
  \bibfield  {author} {\bibinfo {author} {\bibfnamefont {C.}~\bibnamefont
  {Quigg}},\ }\href {https://books.google.com/books?id=kXiWmwEACAAJ} {\emph
  {\bibinfo {title} {{Gauge Theories of the Strong, Weak, and Electromagnetic
  Interactions: Second Edition}}}}\ (\bibinfo  {publisher} {Princeton
  University Press},\ \bibinfo {year} {2013})\BibitemShut {NoStop}%
\bibitem [{\citenamefont
  {Pancharatnam}(1956)}]{Pancharatnam1956GeneralizedPencils}%
  \BibitemOpen
  \bibfield  {author} {\bibinfo {author} {\bibfnamefont {S.}~\bibnamefont
  {Pancharatnam}},\ }\bibfield  {title} {\bibinfo {title} {{Generalized theory
  of interference, and its applications - Part I. Coherent pencils}},\ }\href
  {https://doi.org/10.1007/BF03046050} {\bibfield  {journal} {\bibinfo
  {journal} {Proceedings of the Indian Academy of Sciences - Section A}\
  }\textbf {\bibinfo {volume} {44}},\ \bibinfo {pages} {247} (\bibinfo {year}
  {1956})}\BibitemShut {NoStop}%
\bibitem [{\citenamefont {Berry}(1984)}]{Berry1984QuantalChanges}%
  \BibitemOpen
  \bibfield  {author} {\bibinfo {author} {\bibfnamefont {M.~V.}\ \bibnamefont
  {Berry}},\ }\bibfield  {title} {\bibinfo {title} {{Quantal phase factors
  accompanying adiabatic changes}},\ }\href
  {https://doi.org/10.1098/rspa.1984.0023} {\bibfield  {journal} {\bibinfo
  {journal} {Proceedings of the Royal Society of London. A. Mathematical and
  Physical Sciences}\ }\textbf {\bibinfo {volume} {392}},\ \bibinfo {pages}
  {45} (\bibinfo {year} {1984})}\BibitemShut {NoStop}%
\bibitem [{\citenamefont {Simon}(1983)}]{Simon1983HolonomyPhase}%
  \BibitemOpen
  \bibfield  {author} {\bibinfo {author} {\bibfnamefont {B.}~\bibnamefont
  {Simon}},\ }\bibfield  {title} {\bibinfo {title} {{Holonomy, the quantum
  adiabatic theorem, and Berry's phase}},\ }\href
  {https://doi.org/10.1103/PhysRevLett.51.2167} {\bibfield  {journal} {\bibinfo
   {journal} {Physical Review Letters}\ }\textbf {\bibinfo {volume} {51}},\
  \bibinfo {pages} {2167} (\bibinfo {year} {1983})}\BibitemShut {NoStop}%
\bibitem [{\citenamefont {Martinelli}\ and\ \citenamefont
  {Vavassori}(1990)}]{Martinelli1990ACircuits}%
  \BibitemOpen
  \bibfield  {author} {\bibinfo {author} {\bibfnamefont {M.}~\bibnamefont
  {Martinelli}}\ and\ \bibinfo {author} {\bibfnamefont {P.}~\bibnamefont
  {Vavassori}},\ }\bibfield  {title} {\bibinfo {title} {{A geometric
  (Pancharatnam) phase approach to the polarization and phase control in the
  coherent optics circuits}},\ }\href
  {https://doi.org/10.1016/0030-4018(90)90380-C} {\bibfield  {journal}
  {\bibinfo  {journal} {Optics Communications}\ }\textbf {\bibinfo {volume}
  {80}},\ \bibinfo {pages} {166} (\bibinfo {year} {1990})}\BibitemShut
  {NoStop}%
\bibitem [{\citenamefont
  {Bhandari}(1991{\natexlab{a}})}]{Bhandari1991SU2Phases}%
  \BibitemOpen
  \bibfield  {author} {\bibinfo {author} {\bibfnamefont {R.}~\bibnamefont
  {Bhandari}},\ }\bibfield  {title} {\bibinfo {title} {{SU(2) phase jumps and
  geometric phases}},\ }\href {https://doi.org/10.1016/0375-9601(91)90055-D}
  {\bibfield  {journal} {\bibinfo  {journal} {Physics Letters A}\ }\textbf
  {\bibinfo {volume} {157}},\ \bibinfo {pages} {221} (\bibinfo {year}
  {1991}{\natexlab{a}})}\BibitemShut {NoStop}%
\bibitem [{\citenamefont
  {Bhandari}(1991{\natexlab{b}})}]{Bhandari1991EvolutionDirection}%
  \BibitemOpen
  \bibfield  {author} {\bibinfo {author} {\bibfnamefont {R.}~\bibnamefont
  {Bhandari}},\ }\bibfield  {title} {\bibinfo {title} {{Evolution of light
  beams in polarization and direction}},\ }\href
  {https://doi.org/10.1016/0921-4526(91)90699-F} {\bibfield  {journal}
  {\bibinfo  {journal} {Physica B: Physics of Condensed Matter}\ }\textbf
  {\bibinfo {volume} {175}},\ \bibinfo {pages} {111} (\bibinfo {year}
  {1991}{\natexlab{b}})}\BibitemShut {NoStop}%
\bibitem [{\citenamefont {Tiwari}(1992)}]{Tiwari1992GeometricClassical}%
  \BibitemOpen
  \bibfield  {author} {\bibinfo {author} {\bibfnamefont {S.~C.}\ \bibnamefont
  {Tiwari}},\ }\bibfield  {title} {\bibinfo {title} {{Geometric phase in
  optics: Quantal or classical?}},\ }\href
  {https://doi.org/10.1080/09500349214551101} {\bibfield  {journal} {\bibinfo
  {journal} {Journal of Modern Optics}\ }\textbf {\bibinfo {volume} {39}},\
  \bibinfo {pages} {1097} (\bibinfo {year} {1992})}\BibitemShut {NoStop}%
\bibitem [{\citenamefont {Bliokh}\ \emph {et~al.}(2019)\citenamefont {Bliokh},
  \citenamefont {Alonso},\ and\ \citenamefont
  {Dennis}}]{Bliokh2019GeometricAspects}%
  \BibitemOpen
  \bibfield  {author} {\bibinfo {author} {\bibfnamefont {K.~Y.}\ \bibnamefont
  {Bliokh}}, \bibinfo {author} {\bibfnamefont {M.~A.}\ \bibnamefont {Alonso}},\
  and\ \bibinfo {author} {\bibfnamefont {M.~R.}\ \bibnamefont {Dennis}},\
  }\bibfield  {title} {\bibinfo {title} {{Geometric phases in 2D and 3D
  polarized fields: Geometrical, dynamical, and topological aspects}},\
  }\bibfield  {journal} {\bibinfo  {journal} {Reports on Progress in Physics}\
  }\textbf {\bibinfo {volume} {82}},\ \href
  {https://doi.org/10.1088/1361-6633/ab4415} {10.1088/1361-6633/ab4415}
  (\bibinfo {year} {2019})\BibitemShut {NoStop}%
\bibitem [{\citenamefont {Simon}\ \emph {et~al.}(1988)\citenamefont {Simon},
  \citenamefont {Kimble},\ and\ \citenamefont
  {Sudarshan}}]{Simon1988EvolvingExperiment}%
  \BibitemOpen
  \bibfield  {author} {\bibinfo {author} {\bibfnamefont {R.}~\bibnamefont
  {Simon}}, \bibinfo {author} {\bibfnamefont {H.~J.}\ \bibnamefont {Kimble}},\
  and\ \bibinfo {author} {\bibfnamefont {E.~C.~G.}\ \bibnamefont {Sudarshan}},\
  }\bibfield  {title} {\bibinfo {title} {{Evolving Geometric Phase and Its
  Dynamical Manifestation as a Frequency Shift: An Optical Experiment}},\
  }\href@noop {} {\bibfield  {journal} {\bibinfo  {journal} {Physical Review
  Letters}\ }\textbf {\bibinfo {volume} {61}},\ \bibinfo {pages} {19} (\bibinfo
  {year} {1988})}\BibitemShut {NoStop}%
\bibitem [{\citenamefont {Bhandari}\ and\ \citenamefont
  {Samuel}(1988)}]{Bhandari1988ObservationInterferometer}%
  \BibitemOpen
  \bibfield  {author} {\bibinfo {author} {\bibfnamefont {R.}~\bibnamefont
  {Bhandari}}\ and\ \bibinfo {author} {\bibfnamefont {J.}~\bibnamefont
  {Samuel}},\ }\bibfield  {title} {\bibinfo {title} {{Observation of
  topological phase by use of a laser interferometer}},\ }\href
  {https://doi.org/10.1103/PhysRevLett.60.1211} {\bibfield  {journal} {\bibinfo
   {journal} {Physical Review Letters}\ }\textbf {\bibinfo {volume} {60}},\
  \bibinfo {pages} {1211} (\bibinfo {year} {1988})}\BibitemShut {NoStop}%
\bibitem [{\citenamefont {Kwiat}\ and\ \citenamefont
  {Chiao}(1991)}]{Kwiat1991ObservationPhoton}%
  \BibitemOpen
  \bibfield  {author} {\bibinfo {author} {\bibfnamefont {P.~G.}\ \bibnamefont
  {Kwiat}}\ and\ \bibinfo {author} {\bibfnamefont {R.~Y.}\ \bibnamefont
  {Chiao}},\ }\bibfield  {title} {\bibinfo {title} {{Observation of a
  Nonclassical Berry's Phase for the Photon}},\ }\href
  {https://doi.org/10.1038/2071238d0} {\bibfield  {journal} {\bibinfo
  {journal} {Physical Review Letters}\ }\textbf {\bibinfo {volume} {66}},\
  \bibinfo {pages} {588} (\bibinfo {year} {1991})}\BibitemShut {NoStop}%
\bibitem [{\citenamefont {Ort{\'{i}}z}\ \emph {et~al.}(2014)\citenamefont
  {Ort{\'{i}}z}, \citenamefont {Yugra}, \citenamefont {Rosario}, \citenamefont
  {Sihuincha}, \citenamefont {Loredo}, \citenamefont {Andr{\'{e}}s},\ and\
  \citenamefont {De~Zela}}]{Ortiz2014PolarimetricPhases}%
  \BibitemOpen
  \bibfield  {author} {\bibinfo {author} {\bibfnamefont {O.}~\bibnamefont
  {Ort{\'{i}}z}}, \bibinfo {author} {\bibfnamefont {Y.}~\bibnamefont {Yugra}},
  \bibinfo {author} {\bibfnamefont {A.}~\bibnamefont {Rosario}}, \bibinfo
  {author} {\bibfnamefont {J.~C.}\ \bibnamefont {Sihuincha}}, \bibinfo {author}
  {\bibfnamefont {J.~C.}\ \bibnamefont {Loredo}}, \bibinfo {author}
  {\bibfnamefont {M.~V.}\ \bibnamefont {Andr{\'{e}}s}},\ and\ \bibinfo {author}
  {\bibfnamefont {F.}~\bibnamefont {De~Zela}},\ }\bibfield  {title} {\bibinfo
  {title} {{Polarimetric measurements of single-photon geometric phases}},\
  }\href {https://doi.org/10.1103/PhysRevA.89.012124} {\bibfield  {journal}
  {\bibinfo  {journal} {Physical Review A}\ }\textbf {\bibinfo {volume} {89}},\
  \bibinfo {pages} {012124} (\bibinfo {year} {2014})}\BibitemShut {NoStop}%
\bibitem [{\citenamefont {Maji}\ \emph {et~al.}(2019)\citenamefont {Maji},
  \citenamefont {Jacob},\ and\ \citenamefont
  {Brundavanam}}]{Maji2019GeometricBeams}%
  \BibitemOpen
  \bibfield  {author} {\bibinfo {author} {\bibfnamefont {S.}~\bibnamefont
  {Maji}}, \bibinfo {author} {\bibfnamefont {P.}~\bibnamefont {Jacob}},\ and\
  \bibinfo {author} {\bibfnamefont {M.~M.}\ \bibnamefont {Brundavanam}},\
  }\bibfield  {title} {\bibinfo {title} {{Geometric Phase and
  Intensity-Controlled Extrinsic Orbital Angular Momentum of Off-Axis Vortex
  Beams}},\ }\href {https://doi.org/10.1103/PhysRevApplied.12.054053}
  {\bibfield  {journal} {\bibinfo  {journal} {Physical Review Applied}\
  }\textbf {\bibinfo {volume} {12}},\ \bibinfo {pages} {1} (\bibinfo {year}
  {2019})}\BibitemShut {NoStop}%
\bibitem [{\citenamefont {Karnieli}\ \emph {et~al.}(2019)\citenamefont
  {Karnieli}, \citenamefont {Trajtenberg-Mills}, \citenamefont {Di~Domenico},\
  and\ \citenamefont {Arie}}]{Karnieli2019ExperimentalConversion}%
  \BibitemOpen
  \bibfield  {author} {\bibinfo {author} {\bibfnamefont {A.}~\bibnamefont
  {Karnieli}}, \bibinfo {author} {\bibfnamefont {S.}~\bibnamefont
  {Trajtenberg-Mills}}, \bibinfo {author} {\bibfnamefont {G.}~\bibnamefont
  {Di~Domenico}},\ and\ \bibinfo {author} {\bibfnamefont {A.}~\bibnamefont
  {Arie}},\ }\bibfield  {title} {\bibinfo {title} {{Experimental observation of
  the geometric phase in nonlinear frequency conversion}},\ }\href
  {https://doi.org/10.1364/optica.6.001401} {\bibfield  {journal} {\bibinfo
  {journal} {Optica}\ }\textbf {\bibinfo {volume} {6}},\ \bibinfo {pages}
  {1401} (\bibinfo {year} {2019})}\BibitemShut {NoStop}%
\bibitem [{\citenamefont {Hannonen}\ \emph {et~al.}(2020)\citenamefont
  {Hannonen}, \citenamefont {Partanen}, \citenamefont {Leinonen}, \citenamefont
  {Heikkinen}, \citenamefont {Hakala}, \citenamefont {Friberg},\ and\
  \citenamefont {Set{\"{a}}l{\"{a}}}}]{Hannonen2020MeasurementInterference}%
  \BibitemOpen
  \bibfield  {author} {\bibinfo {author} {\bibfnamefont {A.}~\bibnamefont
  {Hannonen}}, \bibinfo {author} {\bibfnamefont {H.}~\bibnamefont {Partanen}},
  \bibinfo {author} {\bibfnamefont {A.}~\bibnamefont {Leinonen}}, \bibinfo
  {author} {\bibfnamefont {J.}~\bibnamefont {Heikkinen}}, \bibinfo {author}
  {\bibfnamefont {T.~K.}\ \bibnamefont {Hakala}}, \bibinfo {author}
  {\bibfnamefont {A.~T.}\ \bibnamefont {Friberg}},\ and\ \bibinfo {author}
  {\bibfnamefont {T.}~\bibnamefont {Set{\"{a}}l{\"{a}}}},\ }\bibfield  {title}
  {\bibinfo {title} {{Measurement of the Pancharatnam–Berry phase in two-beam
  interference}},\ }\href {https://doi.org/10.1364/optica.401993} {\bibfield
  {journal} {\bibinfo  {journal} {Optica}\ }\textbf {\bibinfo {volume} {7}},\
  \bibinfo {pages} {1435} (\bibinfo {year} {2020})}\BibitemShut {NoStop}%
\bibitem [{\citenamefont {van Enk}(1993)}]{vanEnk1993GeometricTransfer}%
  \BibitemOpen
  \bibfield  {author} {\bibinfo {author} {\bibfnamefont {S.~J.}\ \bibnamefont
  {van Enk}},\ }\bibfield  {title} {\bibinfo {title} {{Geometric phase,
  transformations of gaussian light beams and angular momentum transfer}},\
  }\href {https://doi.org/10.1016/0030-4018(93)90472-H} {\bibfield  {journal}
  {\bibinfo  {journal} {Optics Communications}\ }\textbf {\bibinfo {volume}
  {102}},\ \bibinfo {pages} {59} (\bibinfo {year} {1993})}\BibitemShut
  {NoStop}%
\bibitem [{\citenamefont {Padgett}\ and\ \citenamefont
  {Courtial}(1999)}]{SphereOfModes}%
  \BibitemOpen
  \bibfield  {author} {\bibinfo {author} {\bibfnamefont {M.~J.}\ \bibnamefont
  {Padgett}}\ and\ \bibinfo {author} {\bibfnamefont {J.}~\bibnamefont
  {Courtial}},\ }\bibfield  {title} {\bibinfo {title} {{Poincare-sphere
  equivalent for light beams containing orbital angular momentum}},\
  }\href@noop {} {\bibfield  {journal} {\bibinfo  {journal} {Optics Letters}\
  }\textbf {\bibinfo {volume} {24}},\ \bibinfo {pages} {430} (\bibinfo {year}
  {1999})}\BibitemShut {NoStop}%
\bibitem [{\citenamefont {Courtial}\ \emph {et~al.}(1998)\citenamefont
  {Courtial}, \citenamefont {Dholakia}, \citenamefont {Robertson},
  \citenamefont {Allen},\ and\ \citenamefont
  {Padgett}}]{Courtial1998MeasurementMomentum}%
  \BibitemOpen
  \bibfield  {author} {\bibinfo {author} {\bibfnamefont {J.}~\bibnamefont
  {Courtial}}, \bibinfo {author} {\bibfnamefont {K.}~\bibnamefont {Dholakia}},
  \bibinfo {author} {\bibfnamefont {D.~A.}\ \bibnamefont {Robertson}}, \bibinfo
  {author} {\bibfnamefont {L.}~\bibnamefont {Allen}},\ and\ \bibinfo {author}
  {\bibfnamefont {M.~J.}\ \bibnamefont {Padgett}},\ }\bibfield  {title}
  {\bibinfo {title} {{Measurement of the Rotational Frequency Shift Imparted to
  a Rotating Light Beam Possessing Orbital Angular Momentum}},\ }\href
  {https://doi.org/10.1007/978-3-030-20381-8{\_}3} {\bibfield  {journal}
  {\bibinfo  {journal} {Physical Review Letters}\ }\textbf {\bibinfo {volume}
  {80}},\ \bibinfo {pages} {3217} (\bibinfo {year} {1998})}\BibitemShut
  {NoStop}%
\bibitem [{\citenamefont {Galvez}\ \emph {et~al.}(2003)\citenamefont {Galvez},
  \citenamefont {Crawford}, \citenamefont {Sztul}, \citenamefont {Pysher},
  \citenamefont {Haglin},\ and\ \citenamefont
  {Williams}}]{Galvez2003GeometricMomentum}%
  \BibitemOpen
  \bibfield  {author} {\bibinfo {author} {\bibfnamefont {E.~J.}\ \bibnamefont
  {Galvez}}, \bibinfo {author} {\bibfnamefont {P.~R.}\ \bibnamefont
  {Crawford}}, \bibinfo {author} {\bibfnamefont {H.~I.}\ \bibnamefont {Sztul}},
  \bibinfo {author} {\bibfnamefont {M.~J.}\ \bibnamefont {Pysher}}, \bibinfo
  {author} {\bibfnamefont {P.~J.}\ \bibnamefont {Haglin}},\ and\ \bibinfo
  {author} {\bibfnamefont {R.~E.}\ \bibnamefont {Williams}},\ }\bibfield
  {title} {\bibinfo {title} {{Geometric Phase Associated with Mode
  Transformations of Optical Beams Bearing Orbital Angular Momentum}},\ }\href
  {https://doi.org/10.1103/PhysRevLett.90.203901} {\bibfield  {journal}
  {\bibinfo  {journal} {Physical Review Letters}\ }\textbf {\bibinfo {volume}
  {90}},\ \bibinfo {pages} {4} (\bibinfo {year} {2003})}\BibitemShut {NoStop}%
\bibitem [{\citenamefont {Habraken}\ and\ \citenamefont
  {Nienhuis}(2010)}]{Habraken2010GeometricOrder}%
  \BibitemOpen
  \bibfield  {author} {\bibinfo {author} {\bibfnamefont {S.~J.}\ \bibnamefont
  {Habraken}}\ and\ \bibinfo {author} {\bibfnamefont {G.}~\bibnamefont
  {Nienhuis}},\ }\bibfield  {title} {\bibinfo {title} {{Geometric phases in
  astigmatic optical modes of arbitrary order}},\ }\bibfield  {journal}
  {\bibinfo  {journal} {Journal of Mathematical Physics}\ }\textbf {\bibinfo
  {volume} {51}},\ \href {https://doi.org/10.1063/1.3456078}
  {10.1063/1.3456078} (\bibinfo {year} {2010})\BibitemShut {NoStop}%
\bibitem [{\citenamefont {Alonso}\ and\ \citenamefont
  {Dennis}(2017)}]{Alonso2017Ray-opticalBeams}%
  \BibitemOpen
  \bibfield  {author} {\bibinfo {author} {\bibfnamefont {M.~A.}\ \bibnamefont
  {Alonso}}\ and\ \bibinfo {author} {\bibfnamefont {M.~R.}\ \bibnamefont
  {Dennis}},\ }\bibfield  {title} {\bibinfo {title} {{Ray-optical
  Poincar{\'{e}} sphere for structured Gaussian beams}},\ }\href
  {https://doi.org/10.1364/optica.4.000476} {\bibfield  {journal} {\bibinfo
  {journal} {Optica}\ }\textbf {\bibinfo {volume} {4}},\ \bibinfo {pages} {476}
  (\bibinfo {year} {2017})}\BibitemShut {NoStop}%
\bibitem [{\citenamefont {Malhotra}\ \emph {et~al.}(2018)\citenamefont
  {Malhotra}, \citenamefont {Guti{\'{e}}rrez-Cuevas}, \citenamefont {Hassett},
  \citenamefont {Dennis}, \citenamefont {Vamivakas},\ and\ \citenamefont
  {Alonso}}]{Malhotra2018MeasuringInterferometry}%
  \BibitemOpen
  \bibfield  {author} {\bibinfo {author} {\bibfnamefont {T.}~\bibnamefont
  {Malhotra}}, \bibinfo {author} {\bibfnamefont {R.}~\bibnamefont
  {Guti{\'{e}}rrez-Cuevas}}, \bibinfo {author} {\bibfnamefont {J.}~\bibnamefont
  {Hassett}}, \bibinfo {author} {\bibfnamefont {M.~R.}\ \bibnamefont {Dennis}},
  \bibinfo {author} {\bibfnamefont {A.~N.}\ \bibnamefont {Vamivakas}},\ and\
  \bibinfo {author} {\bibfnamefont {M.~A.}\ \bibnamefont {Alonso}},\ }\bibfield
   {title} {\bibinfo {title} {{Measuring Geometric Phase without
  Interferometry}},\ }\href {https://doi.org/10.1103/PhysRevLett.120.233602}
  {\bibfield  {journal} {\bibinfo  {journal} {Physical Review Letters}\
  }\textbf {\bibinfo {volume} {120}},\ \bibinfo {pages} {1} (\bibinfo {year}
  {2018})}\BibitemShut {NoStop}%
\bibitem [{\citenamefont {Milione}\ \emph {et~al.}(2012)\citenamefont
  {Milione}, \citenamefont {Evans}, \citenamefont {Nolan},\ and\ \citenamefont
  {Alfano}}]{Milione2012HigherLight}%
  \BibitemOpen
  \bibfield  {author} {\bibinfo {author} {\bibfnamefont {G.}~\bibnamefont
  {Milione}}, \bibinfo {author} {\bibfnamefont {S.}~\bibnamefont {Evans}},
  \bibinfo {author} {\bibfnamefont {D.~A.}\ \bibnamefont {Nolan}},\ and\
  \bibinfo {author} {\bibfnamefont {R.~R.}\ \bibnamefont {Alfano}},\ }\bibfield
   {title} {\bibinfo {title} {{Higher order Pancharatnam-Berry phase and the
  angular momentum of light}},\ }\href
  {https://doi.org/10.1103/PhysRevLett.108.190401} {\bibfield  {journal}
  {\bibinfo  {journal} {Physical Review Letters}\ }\textbf {\bibinfo {volume}
  {108}},\ \bibinfo {pages} {1} (\bibinfo {year} {2012})}\BibitemShut {NoStop}%
\bibitem [{\citenamefont {Voitiv}\ \emph {et~al.}(2022)\citenamefont {Voitiv},
  \citenamefont {Lusk},\ and\ \citenamefont
  {Siemens}}]{Voitiv2022TiltedGeodesics}%
  \BibitemOpen
  \bibfield  {author} {\bibinfo {author} {\bibfnamefont {A.~A.}\ \bibnamefont
  {Voitiv}}, \bibinfo {author} {\bibfnamefont {M.~T.}\ \bibnamefont {Lusk}},\
  and\ \bibinfo {author} {\bibfnamefont {M.~E.}\ \bibnamefont {Siemens}},\
  }\bibfield  {title} {\bibinfo {title} {{Tilted Poincar{\'{e}} sphere
  geodesics}},\ }\href {https://doi.org/10.1364/ol.451127} {\bibfield
  {journal} {\bibinfo  {journal} {Optics Letters}\ }\textbf {\bibinfo {volume}
  {47}},\ \bibinfo {pages} {1089} (\bibinfo {year} {2022})}\BibitemShut
  {NoStop}%
\bibitem [{\citenamefont {Cisowski}\ \emph {et~al.}(2022)\citenamefont
  {Cisowski}, \citenamefont {G{\"{o}}tte},\ and\ \citenamefont
  {Franke-Arnold}}]{Cisowski2022Colloquium:Theory}%
  \BibitemOpen
  \bibfield  {author} {\bibinfo {author} {\bibfnamefont {C.}~\bibnamefont
  {Cisowski}}, \bibinfo {author} {\bibfnamefont {J.~B.}\ \bibnamefont
  {G{\"{o}}tte}},\ and\ \bibinfo {author} {\bibfnamefont {S.}~\bibnamefont
  {Franke-Arnold}},\ }\bibfield  {title} {\bibinfo {title} {{Colloquium:
  Geometric phases of light: Insights from fiber bundle theory}},\ }\href
  {https://doi.org/10.1103/RevModPhys.94.031001} {\bibfield  {journal}
  {\bibinfo  {journal} {Reviews of Modern Physics}\ }\textbf {\bibinfo {volume}
  {94}},\ \bibinfo {pages} {31001} (\bibinfo {year} {2022})}\BibitemShut
  {NoStop}%
\bibitem [{\citenamefont {Lusk}\ \emph {et~al.}(2022)\citenamefont {Lusk},
  \citenamefont {Voitiv}, \citenamefont {Zhu},\ and\ \citenamefont
  {Siemens}}]{Lusk2022TheLens}%
  \BibitemOpen
  \bibfield  {author} {\bibinfo {author} {\bibfnamefont {M.~T.}\ \bibnamefont
  {Lusk}}, \bibinfo {author} {\bibfnamefont {A.~A.}\ \bibnamefont {Voitiv}},
  \bibinfo {author} {\bibfnamefont {C.}~\bibnamefont {Zhu}},\ and\ \bibinfo
  {author} {\bibfnamefont {M.~E.}\ \bibnamefont {Siemens}},\ }\bibfield
  {title} {\bibinfo {title} {{The Anatomy of Geometric Phase for an Optical
  Vortex Transiting a Lens}},\ }\href
  {https://doi.org/10.1103/PhysRevA.105.052211} {\bibfield  {journal} {\bibinfo
   {journal} {Physical Review A}\ }\textbf {\bibinfo {volume} {105}},\ \bibinfo
  {pages} {052211} (\bibinfo {year} {2022})}\BibitemShut {NoStop}%
\bibitem [{\citenamefont {Bhandari}(1989)}]{Bhandari1989SynthesisApproach}%
  \BibitemOpen
  \bibfield  {author} {\bibinfo {author} {\bibfnamefont {R.}~\bibnamefont
  {Bhandari}},\ }\bibfield  {title} {\bibinfo {title} {{Synthesis of general
  polarization transformers. A geometric phase approach}},\ }\href
  {https://doi.org/10.1016/0375-9601(89)90747-0} {\bibfield  {journal}
  {\bibinfo  {journal} {Physics Letters A}\ }\textbf {\bibinfo {volume}
  {138}},\ \bibinfo {pages} {469} (\bibinfo {year} {1989})}\BibitemShut
  {NoStop}%
\bibitem [{\citenamefont {Beijersbergen}\ \emph {et~al.}(1993)\citenamefont
  {Beijersbergen}, \citenamefont {Allen}, \citenamefont {van~der Veen},\ and\
  \citenamefont {Woerdman}}]{Beijersbergen1993AstigmaticMomentum}%
  \BibitemOpen
  \bibfield  {author} {\bibinfo {author} {\bibfnamefont {M.~W.}\ \bibnamefont
  {Beijersbergen}}, \bibinfo {author} {\bibfnamefont {L.}~\bibnamefont
  {Allen}}, \bibinfo {author} {\bibfnamefont {H.~E.}\ \bibnamefont {van~der
  Veen}},\ and\ \bibinfo {author} {\bibfnamefont {J.~P.}\ \bibnamefont
  {Woerdman}},\ }\bibfield  {title} {\bibinfo {title} {{Astigmatic laser mode
  converters and transfer of orbital angular momentum}},\ }\href@noop {}
  {\bibfield  {journal} {\bibinfo  {journal} {Optics Communications}\ }\textbf
  {\bibinfo {volume} {96}},\ \bibinfo {pages} {123} (\bibinfo {year}
  {1993})}\BibitemShut {NoStop}%
\bibitem [{\citenamefont {Aharonov}\ and\ \citenamefont
  {Anandan}(1987)}]{Aharonov1987PhaseEvolution}%
  \BibitemOpen
  \bibfield  {author} {\bibinfo {author} {\bibfnamefont {Y.}~\bibnamefont
  {Aharonov}}\ and\ \bibinfo {author} {\bibfnamefont {J.}~\bibnamefont
  {Anandan}},\ }\bibfield  {title} {\bibinfo {title} {{Phase change during a
  cyclic quantum evolution}},\ }\href
  {https://doi.org/10.1103/PhysRevLett.58.1593} {\bibfield  {journal} {\bibinfo
   {journal} {Physical Review Letters}\ }\textbf {\bibinfo {volume} {58}},\
  \bibinfo {pages} {1593} (\bibinfo {year} {1987})}\BibitemShut {NoStop}%
\bibitem [{\citenamefont {Huang}\ \emph {et~al.}(2012)\citenamefont {Huang},
  \citenamefont {Timmers}, \citenamefont {Roberts}, \citenamefont {Shivaram},\
  and\ \citenamefont {Sandhu}}]{Huang2012ALabs}%
  \BibitemOpen
  \bibfield  {author} {\bibinfo {author} {\bibfnamefont {D.}~\bibnamefont
  {Huang}}, \bibinfo {author} {\bibfnamefont {H.}~\bibnamefont {Timmers}},
  \bibinfo {author} {\bibfnamefont {A.}~\bibnamefont {Roberts}}, \bibinfo
  {author} {\bibfnamefont {N.}~\bibnamefont {Shivaram}},\ and\ \bibinfo
  {author} {\bibfnamefont {A.~S.}\ \bibnamefont {Sandhu}},\ }\bibfield  {title}
  {\bibinfo {title} {{A low-cost spatial light modulator for use in
  undergraduate and graduate optics labs}},\ }\href
  {https://doi.org/10.1119/1.3666834} {\bibfield  {journal} {\bibinfo
  {journal} {American Journal of Physics}\ }\textbf {\bibinfo {volume} {80}},\
  \bibinfo {pages} {211} (\bibinfo {year} {2012})}\BibitemShut {NoStop}%
\bibitem [{\citenamefont {Andersen}\ \emph {et~al.}(2019)\citenamefont
  {Andersen}, \citenamefont {Alperin}, \citenamefont {Voitiv}, \citenamefont
  {Holtzmann}, \citenamefont {Gopinath},\ and\ \citenamefont
  {Siemens}}]{Andersen2019CharacterizingHolography}%
  \BibitemOpen
  \bibfield  {author} {\bibinfo {author} {\bibfnamefont {J.}~\bibnamefont
  {Andersen}}, \bibinfo {author} {\bibfnamefont {S.}~\bibnamefont {Alperin}},
  \bibinfo {author} {\bibfnamefont {A.}~\bibnamefont {Voitiv}}, \bibinfo
  {author} {\bibfnamefont {W.}~\bibnamefont {Holtzmann}}, \bibinfo {author}
  {\bibfnamefont {J.}~\bibnamefont {Gopinath}},\ and\ \bibinfo {author}
  {\bibfnamefont {M.}~\bibnamefont {Siemens}},\ }\bibfield  {title} {\bibinfo
  {title} {{Characterizing vortex beams from a spatial light modulator with
  collinear phase-shifting holography}},\ }\href@noop {} {\bibfield  {journal}
  {\bibinfo  {journal} {Applied Optics}\ }\textbf {\bibinfo {volume} {58}}
  (\bibinfo {year} {2019})}\BibitemShut {NoStop}%
\bibitem [{\citenamefont {Andersen}\ \emph {et~al.}(2021)\citenamefont
  {Andersen}, \citenamefont {Voitiv}, \citenamefont {Siemens},\ and\
  \citenamefont {Lusk}}]{Andersen2021HydrodynamicsFluids}%
  \BibitemOpen
  \bibfield  {author} {\bibinfo {author} {\bibfnamefont {J.~M.}\ \bibnamefont
  {Andersen}}, \bibinfo {author} {\bibfnamefont {A.~A.}\ \bibnamefont
  {Voitiv}}, \bibinfo {author} {\bibfnamefont {M.~E.}\ \bibnamefont
  {Siemens}},\ and\ \bibinfo {author} {\bibfnamefont {M.~T.}\ \bibnamefont
  {Lusk}},\ }\bibfield  {title} {\bibinfo {title} {{Hydrodynamics of
  noncircular vortices in beams of light and other two-dimensional fluids}},\
  }\href {https://doi.org/10.1103/PhysRevA.104.033520} {\bibfield  {journal}
  {\bibinfo  {journal} {Physical Review A}\ }\textbf {\bibinfo {volume}
  {104}},\ \bibinfo {pages} {1} (\bibinfo {year} {2021})}\BibitemShut {NoStop}%
\end{thebibliography}%

\pagebreak
\widetext
\begin{center}
\textbf{\large Experimental Measurement of Geometric Phase of Non-Geodesic Circles: Supplementary Information}

\bigskip

The following additional details are presented in the order they arise in the manuscript. Citations refer to the same references above.
\end{center}
\setcounter{equation}{0}
\setcounter{figure}{0}
\setcounter{table}{0}
\setcounter{page}{1}
\makeatletter
\renewcommand{\theequation}{S.\arabic{equation}}
\renewcommand{\thefigure}{S.\arabic{figure}}

\section*{1. Vortex Expression used in Experiments}
A natural way to express an optical vortex mode using the spherical coordinates of $(\xi,\theta)$ of Fig. 1 and in terms of a Hermite-Gaussian (HG) basis is:
\begin{equation} \label{natural}
    \psi_A = \cos{ \frac{\theta}{2} } \, e^{i \xi} \mathrm{HG}_{10} + \sin{ \frac{\theta}{2} } \, e^{-i \xi} \mathrm{HG}_{01},
\end{equation}
where $\mathrm{HG}_{10}$ and $\mathrm{HG}_{01}$ correspond with HG modes on the $\mathcal{I}_0$ and $\mathcal{I}_{90}$ axes, respectively, of the Sphere of Modes (SoM) of Fig. 1. This expression is similar to Eqn. 3 in the manuscript, but it uses the \textit{fixed} spherical angles rather than the generalized angles $(\delta,\beta)$.

While performing the experiments, however, we chose to employ the perspective of ``virtual tilt'' \cite{Andersen2021HydrodynamicsFluids}. Much like how one often speaks of the physical orientation of diagonally-linear polarized light as being at $45\degree$ rather than calling it $90\degree$ as it is on the Poincar\'e sphere (PS), ``tilt'' allows for a vortex-centric interpretation of orientations. The expression we actually use for all calculations and for generating holograms is:
\begin{equation} \label{tilt}
    \psi_A = \frac{1}{\sqrt{3 + \cos{2 \theta_{\mathrm{tilt}}}}} \frac{4}{\sqrt{\pi} w_0^2} e^{-(x^2 + y^2)/w_0^2} \left[(x + i y \cos{\theta_{\mathrm{tilt}}}) \cos{\xi_{\mathrm{tilt}}} + (y - i x \cos{\theta_{\mathrm{tilt}}}) \sin{\xi_{\mathrm{tilt}}} \right].
\end{equation}
This expression works by taking a vortex on the north pole ($\mathcal{I}_{+}$ on Fig. 1),
stretching it by angle $\theta_{\mathrm{tilt}}$, and rotating it about its centroid by $\xi_{\mathrm{tilt}}$.

The conversion relationships between tilt angles, $(\xi_{\mathrm{tilt}},\theta_{\mathrm{tilt}})$, and spherical coordinates used in Fig. 1, $(\xi,\theta)$, are:
\begin{align}
    \theta =& \arccos{\biggl[ \frac{4\cos{(\theta_{\mathrm{tilt}})} }{3 + \cos{(2\theta_{\mathrm{tilt}})}} \biggl]}, \\
    \xi &= 2 \, \xi_{\mathrm{tilt}}.
\end{align}

For the case of starting on the equator, $\beta = \xi$. So we produce the data of Fig. 3 by increasing the programmed angle $\xi_{\mathrm{tilt}}$ from $0$ to $90\degree$; we also measured $\xi_{\mathrm{tilt}}$ from $90\degree$ to $180\degree$, which simply reproduces the same sized-circles ($\beta$ is only unique up to $180\degree$) in reverse order.

\section*{2. Generalization of Starting States}

It is important to note that the projection measurement of Eqn. 4 in the manuscript is \textit{not} restricted to starting states on the equator. We chose this setting only for clarity of presentation. One may choose any axis on which to center their small circles, as well as any orthogonal basis vectors.

To demonstrate how one would retrieve $\beta$ for a circle with initial and intermediate states arbitrarily oriented on the SoM or PS, we demonstrate how it would be done for the center of circles used in the manuscript. Define a vector that locates the center of the circles, $\vec{\alpha}$. If centered on the $\mathcal{I}_{0}$ axis, $\vec{\alpha} = \hat{x}$. Next, define an initial starting point on the sphere, $\vec{A}$ coinciding with $|A \rangle$ on the SoM or PS. In terms of Cartesian coordinates, $\vec{A}$ lives on the sphere at
\begin{equation}
    \vec{A} = \hat{x} \, \sin{\theta} \cos{\xi} + \hat{y} \, \sin{\theta} \sin{\xi} + \hat{z} \, \cos{\theta}.
\end{equation}

The angular radius of the circle on the sphere, defined between the center point and the arbitrary starting point, is then found from the inner product between the two vectors:
\begin{equation}
\begin{split}
    \beta & = \arccos{\left( \vec{\alpha} \cdot \vec{A} \right)} \\
    & = \arccos{\left( \hat{x} \cdot \vec{A} \right)} \\
    & = \arccos{\left( \sin{\theta} \cos{\xi} \right)}.
\end{split}
\end{equation}
Again, we used here a specific center point of $\hat{x}$ for illustration. The purpose of showing this generalization is to show that $\beta$ can be determined for any initial point belonging to any small circle, and so Eqns. 7 can always be determined analytically and measured experimentally.

\section*{3. Step-by-Step Experimental Example: Vortices}

Firstly, we provide more details comprising the schematic of Fig. 2 (a) in the manuscript. A $\lambda = 532$ nm, collimated, single-mode Gaussian passes through a hologram on a transmission spatial light modulator (SLM), from which the first diffracted order is imaged onto a series of $\pi$-converters (labelled ``$\pi$''). These lenses are locked in an orientation such that they focus in the $y$-axis of the beam. The SLM is controlled by an Epson 83H projector \cite{Huang2012ALabs}, and an Andor Zyla sCMOS detector captures transverse images of the vortex beams. For these measurements, we only measure the beam before the first $\pi$-converter (initial state) and then after (intermediate state), which makes a half-circle trajectory. Going through the second $\pi$-converter, the final state of the beam coincides with the initial state after making a complete circle on the SoM.

To illustrate the process of measuring the data points of Fig. 3 (a), we present one detailed example of extracting the three different phases experimentally (the next section has similar details for polarization using one example data set).

To construct an initial vortex state $\psi_A$---corresponding to $|A \rangle$ at point $A$ on the SoM---with an SLM as depicted in Fig. 2 (a), we program a hologram of the form:
\begin{equation}
    \mathrm{Hologram}(x,y) = \mathrm{abs}\left[e^{i \, \mathrm{arg}[\psi_A]} + e^{i k_{\mathrm{g}} x}\right] \times \frac{\mathrm{abs}(\psi_A)}{\mathrm{max}[\mathrm{abs}(\psi_A)]},
\end{equation}
using the vortex definition $\psi_A$ of Eqn. \ref{tilt}. 

\begin{figure}[h!]
\centering
\includegraphics[width=\linewidth]{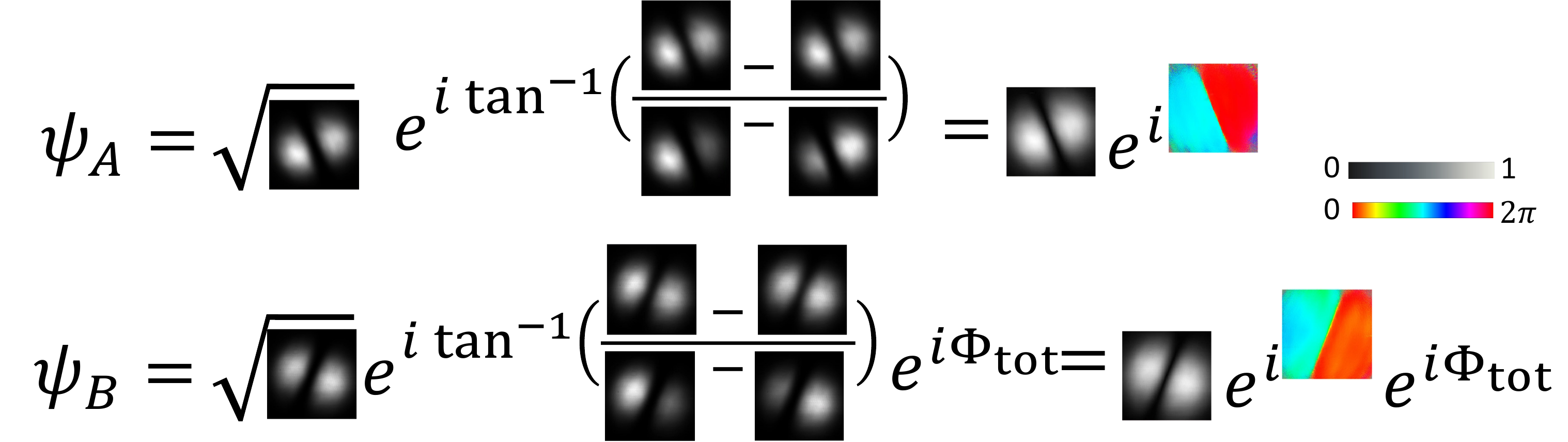}
\caption{Example experimental field measurements for initial vortex condition $\psi_A$ at $(\xi=40\degree,\theta=90\degree)$ and intermediate condition $\psi_B$, which has an acquired phase factor of $\Phi_{\mathrm{tot}}$ that was negated by using a reference Gaussian beam that acquires the same phase shift. The geometric phase can be mined out of the overlaps between the two numerical fields presented here, illustrative of all data points of the paper.}
\label{fig:vortexexample}
\end{figure}

After $4f$-imaging the vortex generated at the SLM and before the first $\pi$-converter, we measure state $\psi_A$ at, for example, SoM coordinates $(\xi=40\degree,\theta=90\degree)$. The amplitude (square-root of the measured intensity) and phase (recovered from four phased-stepped inteferograms via the method of phase-shifting digital holography \cite{Andersen2019CharacterizingHolography}) are combined to produce the numerical complex field of the initial state. This is illustrated in Fig. \ref{fig:vortexexample}, including also intermediate state $\psi_B$ after the first $\pi$-converter, corresponding to $|B \rangle$ at point $B$ on the SoM. 

The Gaussian reference beam used to measure the phase of $\psi_B$ acquires the same total phase of $\Phi_{\mathrm{tot}} = -\pi/2$ after transit through the $\pi$-converter \cite{Beijersbergen1993AstigmaticMomentum}. Therefore, since the phase of $\psi_B$ is measured relative to this Gaussian, this total phase is not actually measured. To remedy this, we include an extra factor of $e^{-i \pi /2}$ on the measured intermediate fields based on this long-established mode-converter theory \cite{Beijersbergen1993AstigmaticMomentum}; this is depicted in the schematic expression of $\psi_B$ in Fig. \ref{fig:vortexexample}.

The first step to use these measurements to calculate the total, dynamic, and geometric phases is to use these numerical fields, as produced in Fig. \ref{fig:vortexexample}, to find the overlap between the states. This is done as shown below in Eqn. \ref{overlap}, where we integrate over all of the camera pixels that contain the beam:
\begin{equation} \label{overlap}
    \begin{split}
        \mathrm{``overlap''} = \int_{\mathrm{all \, pixels}} \psi_A ^* \psi_B \, dx \, dy.
    \end{split}
\end{equation}
This is the calculation of $\langle A | B \rangle$, Eqn. 4 of the manuscript.

For the data shown in Fig. \ref{fig:vortexexample}, ``overlap''$= 0.06 - 0.75 i$, for one set of the five sets of data for that $\beta$ value. Once ``overlap'' is calculated like this, we find the three respective phases by:

\begin{enumerate}
    
    \item \textbf{Total Phase.} For this particular initial condition of $\beta = 40\degree$:
    \begin{equation}
        \Phi_{\mathrm{tot}}|_{\beta=40\degree} = \mathrm{arg} \left( \mathrm{overlap} \right) \times \mathrm{sgn} \left( \cos{(40 \degree)} \right).
    \end{equation}
    Multiplying by $\mathrm{sgn} \left( \cos{\beta} \right)$ ensures that the total phase remains constantly \textit{negative} for any value of $\beta$, to be consistent with the solid-angle definition of geometric phase for partial-circle arcs. (Alternatively, one may simply take modulo $2\pi$ at the end for the full-circle results---which is what we've invoked to do this.)
    
    For the same example data and same ``overlap''$= 0.06 - 0.75 i$ above, this results in $\Phi_{\mathrm{tot}}|_{\beta=40\degree} = \mathrm{arg} \left(0.06 - 0.75 i \right) \times (1) = -1.49$ radians, approximately the expected value of $-\pi/2$. We double this value to get the phase of the complete circle, $-2.98$ radians.
    
    \item \textbf{Dynamic Phase.} Once total phase is found, we can calculate the dynamic phase as shown in Eqn. 7b. Since we have the total phase from the step above, we simply apply Eqn. 6.
    
    \begin{equation}
        \Phi_{\mathrm{dyn}}|_{\beta=40\degree} = \Phi_{\mathrm{tot}}|_{\beta=40\degree} \times -\mathrm{Im} \left( \mathrm{overlap} \right).
    \end{equation}
    
    For the same ``overlap''$= 0.06 - 0.75 i$ above, the dynamic phase is found to be: 
    \begin{equation*}
        \begin{split}
            \Phi_{\mathrm{dyn}}|_{\beta=40\degree} & = (-2.98 \, \mathrm{rad}) \times -\mathrm{Im} \left((0.06 - 0.75 i) \right) \\
            & = (-2.98 \mathrm{rad}) \times \left( 0.75 \right) \\
            & = -2.2 \, \mathrm{rad}.
        \end{split}
    \end{equation*}

    \item \textbf{Geometric Phase.} The geometric phase is then found by:
    \begin{equation}
        \Phi_{\mathrm{geo}}|_{\beta=40\degree} = \Phi_{\mathrm{tot}}|_{\beta=40\degree} - \Phi_{\mathrm{dyn}}|_{\beta=40\degree}.
    \end{equation}
    
    So for the example data shown, $\Phi_{\mathrm{geo}}|_{\beta=40\degree} = -2.98 \, \mathrm{rad} - (-2.2 \, \mathrm{rad}) = -0.78 \, \mathrm{rad}$, for the complete circle.
\end{enumerate}

All three steps are repeated identically for all 5 measurements for a given $\beta$ angle, and then repeated identically for all $\beta$ angles as shown in Fig. 3 (a). We emphasize that we never invoke explicitly a $\cos{\beta}$ relationship on the experimental data---the plotted trend in Fig. 3 (a) arises directly from the numerical overlaps of the measured fields detailed above. Lastly, the size of the error bars (for both experiments) are limited by general optical stability of the set-up: floor/table vibrations and slight air turbulence that cause ``beam jitters'' between one data acquisition and the next.

\section*{4. Step-by-Step Experimental Example: Polarization}

Firstly, we provide more details comprising the schematic of Fig. 2 (b) in the manuscript. A $\lambda = 633$ nm, collimated, single-mode Gaussian is locked to a horizontal polarization with the first linear polarizer (``Lin. Pol.''). A half-waveplate ($\frac{\lambda}{2}$) is used to tune the outgoing intensity of the light to prevent over-saturation on the camera. The beam is sent to a beamsplitter, from which one arm is unmodified to act as reference for total phase measurements. The other arm is sent through another half-waveplate which is mounted in a motorized rotation stage (Thorlabs PRM1Z8 driven by Thorlabs KDC101). This rotation changes the orientation of the waveplate, $\eta$, which increases $\beta$ on the PS according to $\beta = 2\eta$.

\begin{figure}[h!]
\centering
\includegraphics[width=\linewidth]{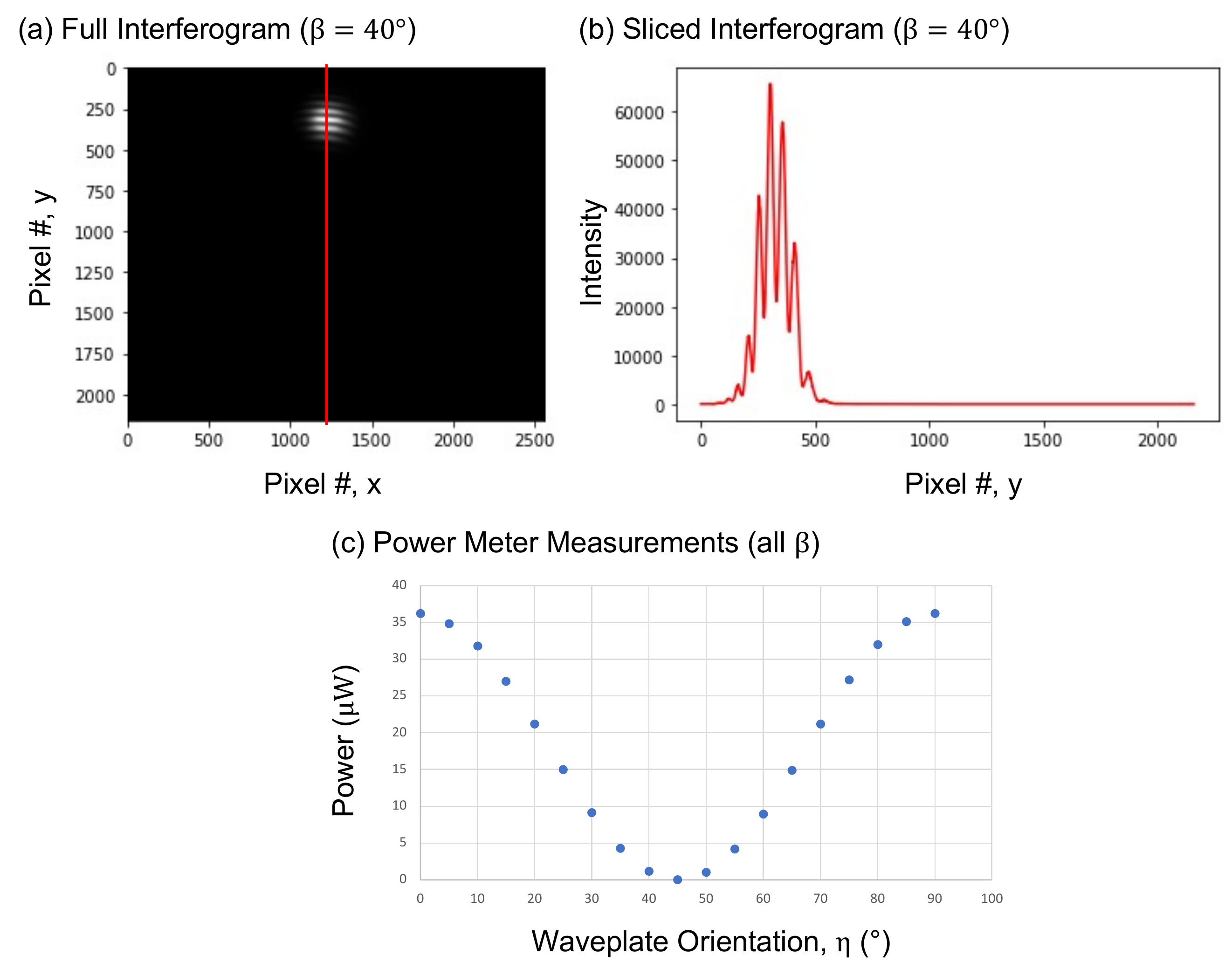}
\caption{\textbf{(a)} Example interference pattern used to measure the total phase associated with a polarization small circle. \textbf{(b)} The corresponding ``slice'' of \textbf{(a)}, depicted with the red line. Shifts in the plotted fringes yield the total phase. \textbf{(c)} Power meter reading of the complete data set, $0\degree \le \beta \le 180\degree$. Malus's law, $\mathrm{I}_{\mathrm{measured}} \propto \cos^2{\left( \eta - 0\degree \right)}$, is clearly evident (the incident polarization is locked at $0\degree$).}
\label{fig:polarizationexample}
\end{figure}

Total phase is measured from the interference between the arms of the interferometer on a camera (Andor Zyla sCMOS), after the beam is reflected back through the half-waveplate to make a complete circle circuit. An example interference pattern is shown in Fig. \ref{fig:polarizationexample} (a), where (b) shows the vertical slice used to measure shifts in the fringes. 25 interference images are taken per each set of data.

For each data set (each $\beta$ value), and for each of the 25 images per set, the total phase is extracted by fitting the slice of Fig. \ref{fig:polarizationexample} (b) to the following model expression:
\begin{equation}
    \mathrm{``model''} = a \, \cos^2{\left(\frac{2\pi}{\lambda_{\mathrm{fringe}}} x + \phi_{\mathrm{shift}}\right) \, e^{-(x-x_0)^2 / w_0^2} + b},
\end{equation}
where the parameters used are: $a$ is the intensity, $b$ is the background level, $\lambda_{\mathrm{fringe}}$ is the width of the fringes, $x_0$ is the center location of the Gaussian-envelope with width $w_0$, and $\phi_{\mathrm{shift}}$ is the sought-after phase shift of the fringes. The average of this phase shift over all 25 images is the desired $\Phi_{\mathrm{tot}}$. We set $\Phi_{\mathrm{tot}}|_{\beta=90\degree} = -\pi$ (the great circle case where there is no dynamic phase), and scale all other total phase measurements to this value.

Once the total phase is measured, the flip mirror is flipped down after the beam makes its first pass through the waveplate (a half-circle is completed on the PS). To measure the dynamic phase, we make a projection measurement on this beam by transmitting it through a linear polarizer locked to the initial polarization of the beam. The result of these measurements, as the waveplate is rotated, is shown in Fig. \ref{fig:polarizationexample} (c) which shows the expected trend of Malus law. This is \textit{nearly} the projection measurement (Eqn. 4 of the manuscript) when we measure the power of the transmitted beam. The polarizer acts on the intensity (rather than the field) of the beam, yielding Malus's Law: $\mathrm{I}_{\mathrm{transmitted}} = \mathrm{I}_{\mathrm{incident}} \, \cos^2{\beta}$. This data in plotted in Fig. \ref{fig:polarizationexample} (c). Thus we modify the power meter measurements to be $\mathrm{A}_{\mathrm{transmitted}} = \mathrm{sgn} \left( \cos{\beta} \right) \sqrt{\mathrm{I}_{\mathrm{transmitted}}}$. The values are then the fractional values of the constant total phase measured previously: $\Phi_{\mathrm{dyn}}(\beta) = \frac{\mathrm{A}(\beta)}{\mathrm{max}\left( \mathrm{A_{\mathrm{all}}} \right)} \times \Phi_{\mathrm{tot}}(\beta)$. From this, $\Phi_{\mathrm{geo}} = \Phi_{\mathrm{tot}} - \Phi_{\mathrm{dyn}}$. 

\end{document}